%% file: main.tex
\documentclass[pra,aps,twocolumn,nopacs,superscriptaddress,nofootinbib,longbibliography]{revtex4-2} 

\usepackage{amsmath}  \usepackage{amssymb}  \usepackage{amsfonts}  \usepackage{bm}  \usepackage{bbm}   \usepackage{braket}  \usepackage{color}  \usepackage{comment}  \usepackage{dcolumn}  \usepackage{enumerate}  \usepackage{epsfig}  \usepackage{gensymb}  \usepackage{graphicx}  \usepackage{indentfirst}  \usepackage{lmodern}  \usepackage{mathrsfs}  \usepackage{mathtools}  \usepackage{psfrag}  \usepackage{pst-all}  \usepackage{soul}  
\usepackage{xcolor}  \usepackage{latexsym}  \usepackage[open]{bookmark}
\usepackage{hyperref}
\usepackage{float} 
\usepackage[T1]{fontenc} 

\include{defs}  
\begin{document}

\title{
Relativistic electron energy-loss spectroscopy in cylindrical waveguides and holes}

\author{Álvaro~Rodríguez~Echarri
}
\email{Corresponding author: arepb15@gmail.com}
\affiliation{Max-Born-Institut, 12489 Berlin, Germany}
\author{Wenhua~Zhao
}
\affiliation{Humboldt-Universität zu Berlin, Institut für Physik, AG Theoretische Optik and Photonik, 12489 Berlin, Germany}
\affiliation{Max-Born-Institut, 12489 Berlin, Germany}
\author{Kurt~Busch
}
\affiliation{Humboldt-Universität zu Berlin, Institut für Physik, AG Theoretische Optik and Photonik, 12489 Berlin, Germany}
\affiliation{Max-Born-Institut, 12489 Berlin, Germany}
\author{F. Javier Garc\'{i}a de Abajo
}
\email{Corresponding author: javier.garciadeabajo@nanophotonics.es}
\affiliation{ICFO-Institut de Ciencies Fotoniques, The Barcelona Institute of Science and Technology, 08860 Castelldefels (Barcelona), Spain}
\affiliation{ICREA-Institucio Catalana de Recerca i Estudis Avancats, Passeig Lluis Companys 23, 08010 Barcelona, Spain}

\begin{abstract}
Swift electrons passing near or through metallic structures have proven to be an excellent tool for studying plasmons and other types of confined optical modes involving collective charge oscillations in the materials hybridized with electromagnetic fields. In this work, we provide a general analytical framework for the simulation of electron energy-loss spectroscopy (EELS) in infinite systems with cylindrical symmetry, such as wires, holes, and optical fibers. While EELS theory is well developed for electrons moving parallel to the direction of translational symmetry, we introduce closed-form analytical solutions for perpendicular electron trajectories. These analytical results are corroborated by comparison to numerical simulations based on a frequency-domain boundary-element method and a discontinuous-Galerkin time-domain finite-element method. Numerical methods further allow us to study termination effects in finite-sized cylindrical objects such as nanorods. The present study of the interaction between free electrons and cylindrically symmetric photonics systems can find application in the analysis of EELS spectra and the design of free-electron--photonic hybrid systems.
\end{abstract}

\maketitle 
\date{\today} 

\section{Introduction}

The study of polaritons --the quanta of collective charge oscillations in material structures hybridized with electromagnetic fields-- drives the field of polaritonics, which is currently fueled by the expectation of promising applications~\cite{M07,paper283}. For instance, plasmon polaritons are able to concentrate light down to subwavelength volumes, thereby considerably enhancing the electromagnetic field at prescribed spatial regions and resonance frequencies \cite{LK00,M06_4}. Upon inelastic decay, the energy carried by plasmon polaritons is partly transferred to electrons in the material, giving rise to localized out-of-equilibrium electron distributions characterized by the presence of hot carriers~\cite{ZHG17}. The spatial and spectral properties of plasmon polaritons can be customized through the choice of material and morphology for applications such as nanoimaging \cite{VWH10}, optical sensing \cite{SAT08}, nonlinear optics~\cite{BBH03}, quantum-state manipulation~\cite{paper327}, and the control of chemical reactions~\cite{BQ14}.

The development of polaritonic applications benefits from the ability to characterize the aforementioned properties. Probing polaritons in nanostructures generally requires an adequate combination of spatial and spectral resolution. While optical spectroscopy provides high energy resolution, spatial resolution is limited by diffraction in far-field techniques, as well as by symmetry selection rules preventing the coupling to dark modes (e.g., highly localized, multipolar excitations). In addition, surface polaritons propagating along planar interfaces feature dispersion relations below the light cone and, therefore, cannot be directly excited via far-field illumination. Instead, tips, prisms, and gratings are commonly used to assist the coupling of freely propagating light and confined polaritons \cite{M07,paper283}, although these methods produce unwanted modifications in the studied specimen.

Electron beams (e-beams) are relatively noninvasive tools that can excite and probe confined polaritons, to which they couple through their associated broadband electromagnetic field \cite{paper149}. Understandably, e-beams have been exploited to characterize spectrally and spatially image polaritons ever since the first experimental demonstration of the existence of surface plasmons \cite{PS1959}, as illustrated by subsequent measurements of plasmon-polariton properties in thin films \cite{CS1975_2} and more complex nanostructures \cite{RB13}. Scanning transmission electron microscopy (STEM) \cite{BJT07,SMM18} currently offers a spatial resolution $\lambda_{e}/{\rm NA}$ in the sub-{\AA}ngström regime thanks to the small de Broglie wavelength of energetic electrons (e.g., $\lambda_{e} \sim 4\,$pm at 100\,keV) combined with numerical apertures ${\rm NA}\sim10^{-2}$ in existing setups. A concurrent energy resolution of a few meV is provided by electron energy-loss spectroscopy (EELS) performed in state-of-the-art STEM instruments \cite{KLD14}, enabling the study of a wide range of polaritons extending from visible and near-infrared plasmons \cite{BKW07,KS14,Y16,WLC17,paper338} to infrared vibrational modes \cite{HRK20,paper412} with atomic resolution.

On the theoretical ground, STEM-EELS can be investigated via analytical theory for simple geometries (e.g., aloof excitation of Mie modes in spherical particles \cite{paper021}, and more recently also for penetrating trajectories \cite{SZR24}), while more complex structures are efficiently simulated through numerical solutions of the Maxwell equations \cite{paper040,MNH11,TSV13,H14_4}. Overall, analytical solutions are favored because they provide deeper insight and permit conducting more exhaustive explorations of different combinations of materials and geometries. However, some simple geometries are still lacking analytical theory, for example, extended metallic cylinders and infinitely extended holes drilled in a homogeneous material. Solutions for such cylindrical systems are only available for e-ebeam trajectories parallel to the axis of symmetry \cite{ZRE1989,PR97,SS01} with and without the inclusion of retardation, an extension of nonretarded theory to incorporate nonlocal effects \cite{AU98,AR08}, and a semi-analytical method for perpendicular trajectories \cite{TSV15}.

In the present work, we provide a self-contained derivation of closed-form analytical expressions for STEM-EELS, including retardation in infinitely extended cylindrical systems both for parallel and perpendicular e-beam trajectories relative to the axis of symmetry. We corroborate our analytical results by comparison to numerical simulations using a boundary-element method (BEM) \cite{paper040} and relate infinitely extended systems to long but finite cylindrical structures via numerical simulations based on a discontinuous-Galerkin time-domain (DGTD) finite-element approach \cite{MNH11,SZR24}. Furthermore, we apply our theory to dielectric cylindrical structures featuring waveguided modes that are analogous to surface plasmon-polaritons in metallic systems. Specifically, we determine the coupling strength between an e-beam to these guided modes. Our analytical approach bears relevance to understanding electron-light interaction in the emerging field of free-electron nanophotonics \cite{SWY15,GTS16,paper372,TDW23}.

\begin{figure*}
    \centering
    \includegraphics[width=1\textwidth]{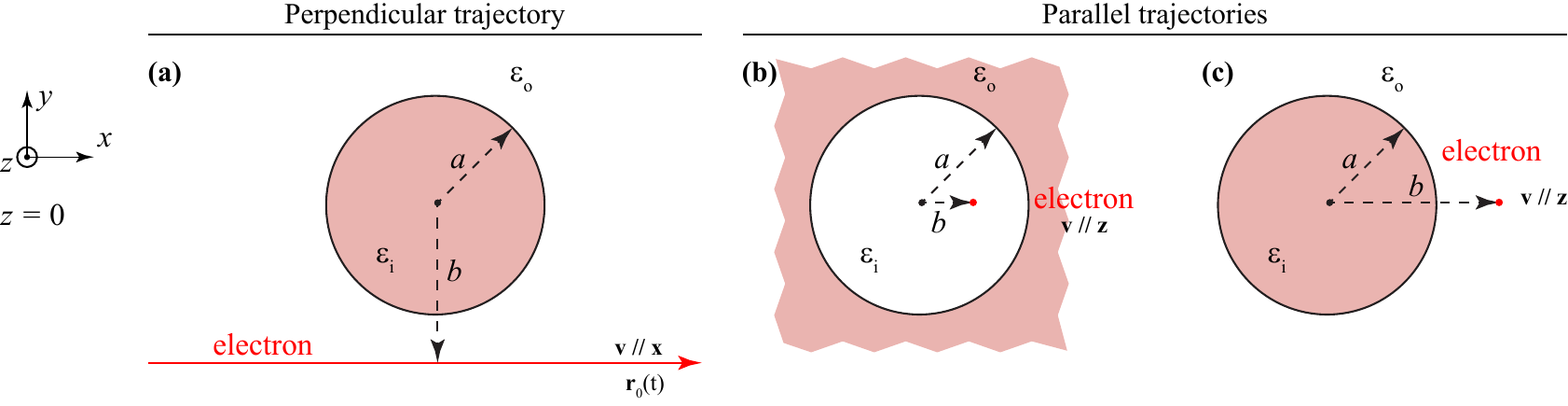}
    \caption{\textbf{Electron energy-loss spectroscopy in cylindrical configurations}. A swift electron (red arrow, dots, and labels) is moving with velocity $\vb$ either perpendicular (a) or parallel (b,c) to the axis of symmetry. We consider external (a,c) and internal (b) trajectories relative to the cylindrical surface of radius $a$, with a distance (impact parameter) $b$ separating the electron trajectory from the axis of symmetry. The media inside and outside the cylinder are described through homogeneous isotropic permittivities $\epso$ and $\epsi$, respectively.
    }
\label{fig:fig1}
\end{figure*}

\section{Theory of electron energy-loss spectroscopy in cylindrical systems}

Consider an electron moving with velocity $\vb$ and interacting with a photonic structure. The electron is described as a moving point charge whose electromagnetic field induces a response in the material. The so-induced electromagnetic fields act back on the electron, producing an overall frictional force that results in electron energy loss. The latter is computed as \cite{paper149}
\begin{align}
    \Delta E = e \int_{-\infty}^\infty dt \, \vb \cdot \Eb^\ind[\rb_e(t),t] = \int_0^\infty d \omega \, \hbar\omega \, \Gamma_\EELS(\omega), \nonumber
\end{align}
where $-e<0$ is the electron charge, $\rb_e(t)$ describes the electron position, $\Eb^\ind[\rb_e(t),t]$ is the self-induced electric field experienced by the electron, and 
\begin{align} \label{Eq:EELS_dt}
    \Gamma_\EELS(\omega)  = \frac{e}{ \pi \hbar \omega} \int_{-\infty}^\infty dt\, \real{\ee^{-\ii \omega t} \, \vb \cdot \Eb^\ind  [\rb_e(t),\omega]}
\end{align}
is the so-called electron energy-loss (EEL) probability per unit of transferred frequency $\omega$. The total energy lost by the electron $\Delta E \lesssim 1$~eV is typically negligible for the electron acceleration voltages used in electron microscopes (few-to-100s~keV). Thus, we can assume that the electron travels along a straight line with a constant velocity vector (nonrecoil approximation). Under these conditions, the longitudinal momentum (along $\vb$) transferred from the electron to the specimen is $\hbar q_z \approx \Delta E/v \ll m_ev$, where $m_e$ is the free electron mass \cite{paper149}.

In what follows, we derive analytical expressions for the EEL probability experienced by swift electrons interacting with individual cylindrical waveguides and holes when the electrons are moving along either parallel or perpendicular trajectories relative to the symmetry axis, as shown in Fig.\ref{fig:fig1}, where $\epsi$ is the permittivity "inside" the cylinder and $\epso$ the "outside" one. The calculation procedure is similar in all scenarios. First, we compute the electric field produced by the electric current distribution $\Jb(\rb,t) = -e \vb \delta[\rb-\rb_e(t)]$ associated with the moving point-like electron. Inside a homogeneous medium $j\in\{\rm i,\rm o\}$, the field is
\begin{align}\label{Eq:E_field_current}
    \Eb_j(\rb,\omega) = \frac{\ii}{\omega \eps_j}  \left(k^2_j + \nabla \otimes \nabla \right) \int d^3\rb' \, \frac{\ee^{\ii k_j |\rb-\rb'|}}{|\rb-\rb'|} \Jb(\rb',\omega),
\end{align}
where $\Jb(\rb,\omega) = \int dt \, \ee^{\ii \omega t} \Jb(\rb,t)$ is the Fourier transform of the current density, and the wavevector $\kb_j=(\Qb_j,q_z)$ is defined to have components
\begin{subequations}
\begin{align}
    k_j &= \sqrt{\eps_j \mu_j (\omega/c)^2}, \label{Eq:kj_definition} \\
    Q_j &= \sqrt{k_j^2-q_z^2+\ii 0^+} \label{Eq:Qj_definition}.
\end{align}
\end{subequations}
Second, we obtain the induced field by imposing the boundary conditions and evaluate it at the position of the electron to calculate the EEL probability through Eq.~\eqref{Eq:EELS_dt}. For simplicity, we consider the electron to be moving in vacuum (e.g., outside a self-standing wire or inside a hole).

Given the symmetry of the systems under consideration, it is convenient to expand the external electric field in terms of cylindrical waves. Inside a homogeneous material $j$ free of external sources and characterized by a dielectric function $\eps_j$ (and unit magnetic permeability), we can label the elementary cylindrical waves with the azimuthal number $m$, the momentum $q_z$ along the axis of symmetry $\zz$, and the polarization $\sigma \in \{s,p\}$. More precisely, following the notation of Ref.~\cite{paper047}, we have
\begin{align}
    \Eb^\xi_{j,q_z ms} (\rb,\omega) = \left[ \frac{\ii m }{Q_j R} \xi_m(Q_j R) \hat{\Rb}-\xi_m'(Q_jR)\hat{\varphi} \right] \ee^{\ii m \varphi} \ee^{\ii q_z z} \nonumber
\end{align}
for $s$ polarization, and
\begin{align}
    \Eb^\xi_{j,q_z mp}(\rb,\omega) = \frac{q_z}{k_j} \bigg[& \ii \xi_m'(Q_j R) \hat{\Rb}-\frac{m}{Q_j R} \xi_m(Q_j R) \hat{\varphi} \nonumber \\
    &+\frac{Q_j}{q_z} \xi_m(Q_j R) \hat{z} \bigg] \ee^{\ii m \varphi} \ee^{\ii q_z z} \nonumber
\end{align}
for $p$ polarization. Here, we take $\xi_m=J_m$ (Bessel functions) and $\xi_m=H_m^{(1)}$ (Hankel functions) for regular and outgoing waves, the prime indicates the derivative with respect to the argument, and we use the notation $\rb=(\Rb,z)$ with $\Rb=(x,y)$. Regular waves are nonsingular at the axis $R=0$, while outgoing waves are not. These waves satisfy the orthogonality relation
\begin{align}
    &\int_0^\infty R dR\, \int_0^{2\pi} d\varphi \, \Eb^J_{j,q_z m\sigma} (\rb) \cdot \left[\Eb^J_{j,q_z'm'\sigma'} (\rb) \right]^* \nonumber  \\ 
    &= 2\pi \delta_{mm'} \delta_{\sigma \sigma'} \frac{1}{q_z} \delta(q_z-q_z'), \nonumber
\end{align}
which we use to obtain analytical expressions for the expansion coefficients of the fields in cylindrical waves by applying the electromagnetic boundary conditions.

\subsection{Perpendicular trajectory}

\begin{figure*}
    \centering
    \includegraphics[width=1\textwidth]{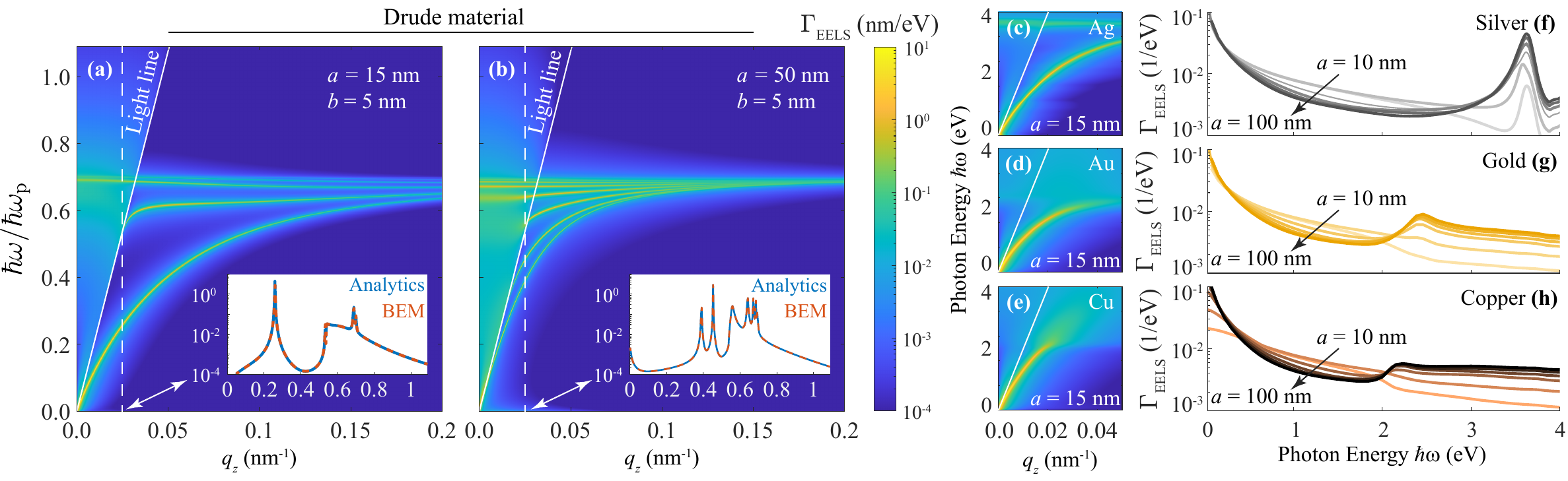}
    \caption{\textbf{EEL probability for electron trajectories perpendicular to infinitely long cylindrical metallic structures}.
    (a,b) Momentum-resolved EEL probability for silver cylinders of radii (a) $a=15$~nm and (b) $a=50$~nm, with the metal described through the Drude model ($\hbar\wp=9.17$~eV and $\hbar\gamma=21$~meV). The e-beam moves in vacuum with an energy of 100~keV (velocity $v\approx0.55c$) and passes 5~nm away from the cylinder surface. The light line (white) is shown for reference. The colormap is shown on a logarithmic scale. The insets show frequency cuts for fixed longitudinal wave vector $q_z=0.025$~nm$^{-1}$, computed by using the analytical Eq.~\eqref{eq:EELS_perp_q} (blue curve) or the numerical BEM (red curve). 
    (c-e) Same as (a), but for cylinders made of silver (Ag), gold (Au), and copper (Cu), calculated by using their measured dielectric functions \cite{JC1972}. The color scale is shared with panels (a,b).
    (f-h) Total EEL probability for cylinders of radii $a = 10,\, 20,\, 40,\, 60,\, 80,\,$ and $100$~nm for the same metals as in (c-e) using 100~keV electrons passing 5~nm away from the metal surface.
    }
\label{fig:EELS_perp}
\end{figure*}

We consider a point-like electron moving along $\xx$ within the host medium of permittivity $\epso$, following the trajectory $\rb_e(t)=(vt,-b,0)$ with $b>0$ denoting the impact parameter (see Fig.~\ref{fig:fig1}a). Such an electron produces a current density
\begin{align}
    \Jb(\rb,\omega) = -e \vv \delta(z) \delta(y+b) \ee^{\ii \omega x / v}  \nonumber
\end{align}
in frequency space. Inserting this expression into Eq.~\eqref{Eq:E_field_current}, we obtain the external electric field
\begin{align}
    \Eb^\ext(\rb,\omega) =-\frac{\ii e}{\omega \epso } (k^2\epso+\nabla \partial_x) \int dx' \frac{\ee^{\ii \ko |\rb-\rb'|}}{|\rb-\rb'|} \ee^{\ii \omega x'/v},  \nonumber
\end{align}
with $\rb'=(x',-b,0)$. We proceed by using the momentum representation of the Coulomb-type interaction 
\begin{align}
    \frac{\ee^{\ii \ko |\rb-\rb'|}}{|\rb-\rb'|} &= \int \frac{d^3\qb}{(2\pi)^3} \frac{4\pi}{q^2-\kko}
    \; \ee^{\ii \qb \cdot (\rb-\rb')} \nonumber\\
    &=\frac{1}{2\pi}\int \frac{dq_x dq_z}{\Delta} \;\ee^{\ii(q_xx+q_zz)-\Delta |y+b|}, \nonumber
\vspace{5pt} 
\end{align}
where the second line is obtained by integrating in the complex $q_y$ plane, and we define $\Delta \equiv \sqrt{(\omega/v)^2+q_z^2-\kko}$. Using this result, the external field reduces to 
\begin{align}
    \Eb^\ext(\rb,\omega) =& -\frac{\ii e}{\omega \epso } (k^2\epso+\nabla\partial_x)  \int \frac{dq_z}{\Delta}\,\ee^{\ii \Qb\cdot\Rb} \ee^{\ii q_z z} \ee^{-\Delta b},  \nonumber
\end{align}
with $\Qb = (\omega/v,\ii\Delta)$. We now invoke the Jocobi-Anger identity
\begin{align}
    \ee^{\ii\Qb\cdot\Rb} = \sum_m \ii^m \ee^{\ii m (\varphi-\varphi_\Qb)} J_m(Q R) \nonumber
\end{align}
to recast the external field into
\begin{align}
    \Eb^\ext(\rb,\omega) 
    = \frac{-e \ko }{\omega \epso } \sum_m \ii^m   \int \frac{dq_z}{\Delta}\, \ee^{-\Delta b}    \ee^{-\ii m \varphi_\Qb} \nonumber \\
     \times \left[ \ko\Eb^J_{s,m} \sin{(\varphi_\Qb)} + \ii q_z \Eb^J_{p,m} \cos{(\varphi_\Qb)} \right], \nonumber
\end{align}
where $\varphi$ and $\varphi_\Qb$ are the azimuthal angles of $\Rb$ and $\Qb$, $Q^2 = \kko-q_z^2$, $\ee^{\pm \ii \varphi_\Qb} = (\omega/v \mp \Delta)/Q$, and we have applied the identity
\begin{align}
    (\kko+\nabla \partial_x) \ee^{\ii q_z z} &\ee^{\ii m \varphi} J_m(\Qo R) =  \nonumber \\
    = \frac{-\ii \kko}{2} &\left[ \Eb^J_{{\rm o},q_z,m+1,s}+\Eb^J_{{\rm o},q_z,m-1,s} \right. \nonumber \\
    +\frac{q_z}{\ko} &\left.\left( \Eb^J_{{\rm o},q_z,m+1,p}-\Eb^J_{{\rm o},q_z,m-1,p} \right)\right] \nonumber
\end{align}
(for a derivation, see Ref.~\cite{paper385}), and the $(\rb,\omega)$ dependence on the fields have been omited for clarity.

We now obtain the induced field by simply multiplying each of the cylindrical-wave components in the external field by the reflection coefficients $t^\out_{\sigma \sigma'}$ of the cylinder for a source outside (see Appendix~\ref{Sec:scattering_outside}). After some algebra, evaluating the induced field at the electron position [i.e., $E^\ind_x(x,-b,0,\omega)$], the EEL probability from Eq.~\eqref{Eq:EELS_dt} reduces to
\begin{align}
    \Gamma_\EELS(\omega) = \int_{-\infty}^\infty dq_z \, \Gamma_\EELS(q_z,\omega)
\label{Gammaqzw}
\end{align}
(an integral over longitudinal wave vectors $q_z$), where the integrand admits the closed-form expression
\begin{widetext}
\begin{align} \label{eq:EELS_perp_q}
     \Gamma_\EELS(q_z,\omega) = 
    \frac{-2e^2}{\pi \hbar c \omega}  \real{ 
    \frac{\ee^{-2\Delta b}    }{\Delta^2 \QQo}\, \sum_m   
    \left\{
    \begin{matrix}
     \Delta  \left[ \ko \Delta t_{m,ss}^\out  -  q_z q t_{m,sp}^\out  \right]  
-   \dfrac{q_z}{\ko}       \left[ \ko \Delta t_{m,ps}^\out  -  q_z q t_{m,pp}^\out  \right] 
    \end{matrix}
    \right\}
    \left( \frac{q+\Delta}{\Qo }\right)^{2m}
    }.
\end{align}
\end{widetext}
In the derivation of Eq.~(\ref{eq:EELS_perp_q}), we have made use of the identity
\begin{align}
    \int_{-\infty}^{\infty} dx\ \ee^{-\ii\omega x / v } H^{(1)}_m(\Qo R) \ee^{\ii m \varphi} \nonumber \\
    =  2 \ii^{m-1} \frac{\ee^{-\Delta b}}{\Delta} (-1)^m \ee^{-\ii m \varphi_\Qb}. \nonumber
\end{align}
Equation~\eqref{eq:EELS_perp_q} provides an analytical solution to the frequency- and wave-vector-resolved EEL probability for an electron passing perpendicularly to an infinite cylindrical wire, thereby extending a previous semi-analytical result \cite{TSV15}.

\begin{figure}
    \centering
    \includegraphics[width=0.48\textwidth]{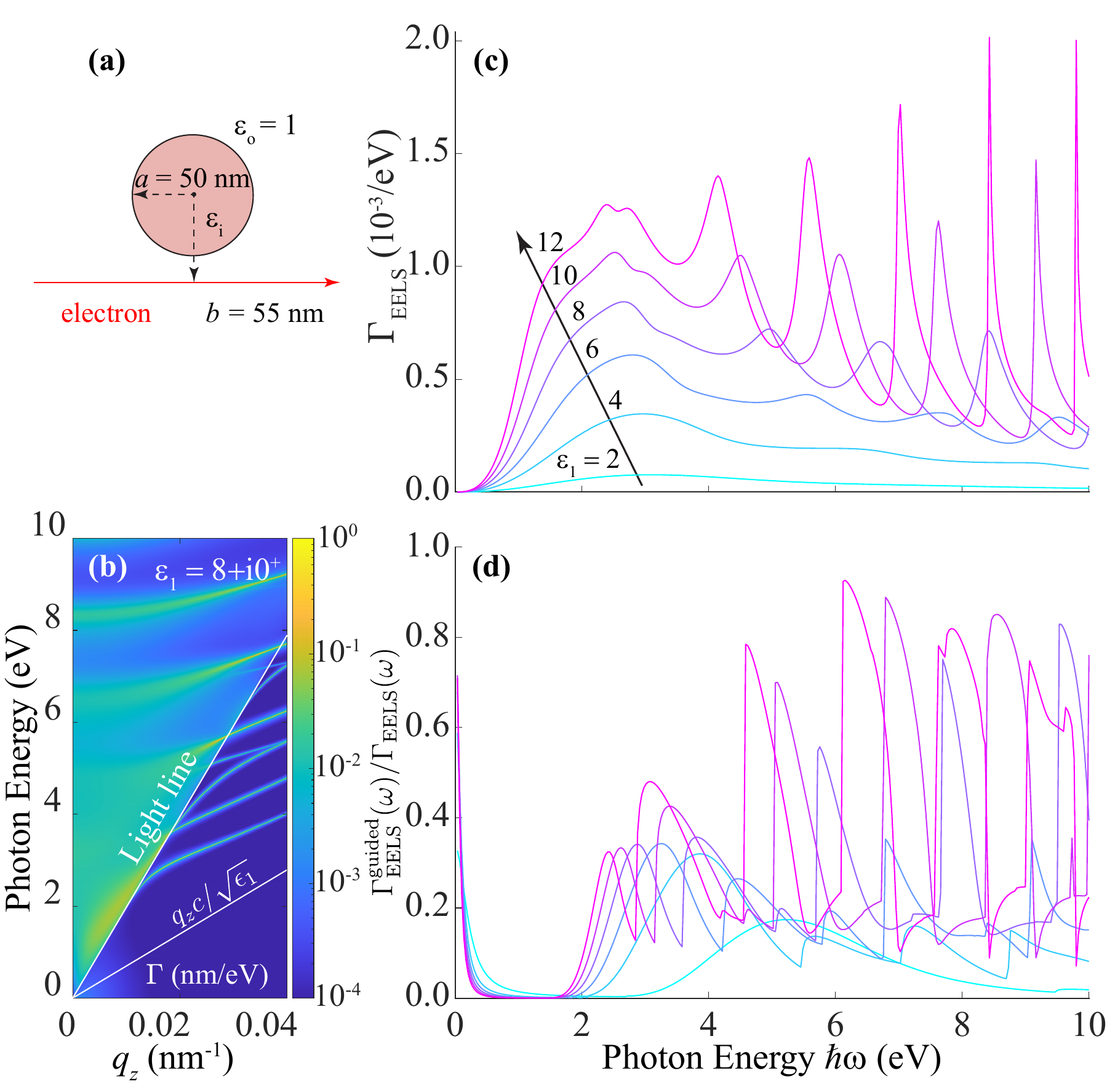}
    \caption{\textbf{EEL probability for optical waveguides under perpendicular e-beam orientation}. 
    (a) Illustration of a cylinder with a radius of 50~nm and permittivity $\epsi$, excited by an electron passing 5~nm away from the surface with a kinetic energy of $100$~keV.
    (b) Momentum-resolved EEL probability $\Gamma_\EELS(q_z,\omega)$ for $\epsi=8+0.001\,\ii$, where a small imaginary part is introduced in the permittivity to help visualize the coupling to guided modes.
    (c) Momentum-integrated EEL probability [Eq.~(\ref{Gammaqzw})] under the same conditions as in panel (a) but for several values of the permittivity ranging from $\eps=2+0.001\,\ii$  to $\eps=12+0.001\,\ii$ (see labels). 
    (d) Ratio between the partial EEL probability $\Gamma^{\rm guided}_\EELS(\omega)$ contributed by guided modes (i.e., the momentum integral outside the light cone, $|q_z|>\omega/c$) and the total EEL probability $\Gamma_\EELS(\omega)$.
    }
\label{fig:fig_dielectric}
\end{figure}

\subsection{Parallel trajectory}

\begin{figure*}
    \centering
    \includegraphics[width=1\textwidth]{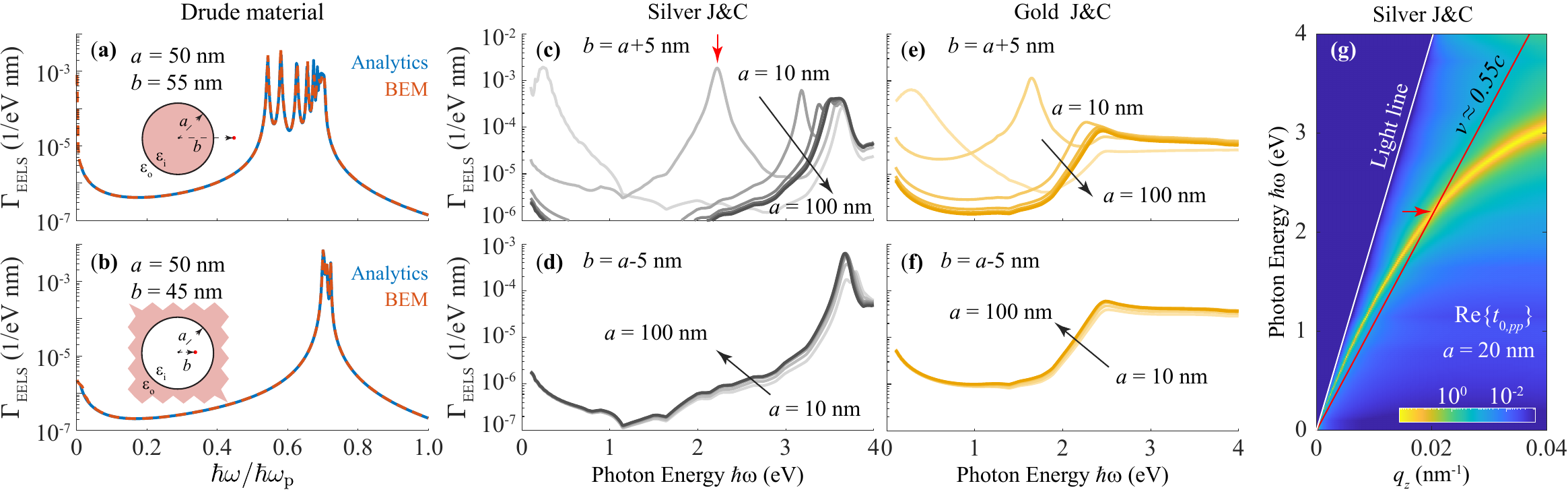}
    \caption{\textbf{EEL probability for electron trajectories parallel to infinitely long cylindrical structures}.
    \textbf{(a,b)} EEL probability for an electron moving parallel to the axis of (a) a silver cylinder and (b) a hole in silver. The electron moves at a distance $|b-a|=5$~nm from the surface. The probability is calculated by using the corresponding analytical formulas (blue curve) or the numerical BEM (red curve). The metal is modelled by a Drude permittivity with $\hbar\wp=9.17$~eV and $\hbar\gamma=21$~meV. The electron energy is $100$~keV.
    (c,d) EEL probability for (c) a silver cylinder and (c) a hole in silver with radii $a = 10,\, 20,\, 40,\, 60,\, 80,\,$ and $100$~nm, and a fixed electron-surface distance $|b-a|=5$~nm, with the metal described by using the dielectric function measured by Johnson and Christy (J\&C) \cite{JC1972}.
    (e,f) Same as panels (c,d), but for gold. 
    (g) Energy-momentum map of the transmission coefficient $\real{t^\out_{0,pp}}$ for a silver cylinder of radius $a=20$~nm. The light (white) and electron (red, $v\approx0.55c$) lines are shown for reference. The red arrow indicates the crossing point of the fundamental surface plasmon resonance and the electron line, which is the same energy position as indicated by the red arrow in panel (c).
    }
\label{fig:EELS_parallel}
\end{figure*}

For an electron trajectory parallel to the axis of symmetry of a cylinder, writing the trajectory as $\rb_e(t)=(b,0,vt)$, the external current density is 
\begin{align}
    \Jb(\rb,\omega) = -e \vv \delta(x-b) \delta(y) \ee^{\ii \omega z / v}, \nonumber
\end{align}
and the field obtained from Eq.~\eqref{Eq:E_field_current} when the electron moves in a homogeneous medium $j$ becomes
\begin{align}
    \Eb^\ext(\rb,\omega) = \frac{-e \ii }{\omega \eps_j } (k^2\eps_j+\nabla \otimes \nabla) \int dz' \frac{\ee^{\ii k_j |\rb-\rb'|}}{|\rb-\rb'|} \ee^{\ii \omega z'/v } \zz, \nonumber
\end{align}
where $\rb'= (b,0,z')$ and $q = \omega/v$. Using the identity 
\begin{align}
    \int dz' \frac{\ee^{\ii k_j |\rb-\rb'|}}{|\rb-\rb'|} \ee^{\ii q z' } = \ii \pi \ee^{\ii q z} H^{(1)}_0(Q_j|\Rb-\Rb'|),
    \nonumber
\end{align}
with $k_j$ and $Q_j$ defined in Eqs.~\eqref{Eq:kj_definition} and \eqref{Eq:Qj_definition}, the external field reduces to
\begin{align}
    \Eb^\ext(\rb,\omega) = \frac{ e \pi }{\omega \eps_j } (k^2\eps_j+\nabla \otimes \nabla)  \ee^{\ii q z} H_0(Q_j|\Rb-\Rb'|) \zz. \nonumber
\end{align}
We now apply Graf's theorem (Eq.~9.1.79 in Ref.~\citenum{AS1972}) to separate the $R$ and $R'$ dependence. More precisely, we use
\begin{align}
    H_0(\Qi|\Rb-\Rb'|) = \sum_m H_m(\Qi R ) J_m(\Qi R') \ee^{\ii m (\varphi-\varphi')} \nonumber
\end{align}
for $a=R > R' = b$ and $\varphi'=0$ (hole configuration with the electron in medium $j=\rm i$ \emph{inside} the cylindrical surface); and 
\begin{align}
    H_0(\Qo|\Rb-\Rb'|) = \sum_m H_m(\Qo R' )    J_m(\Qo R) \ee^{\ii m (\varphi-\varphi')} \nonumber
\end{align}
for $a=R < R'=b$ and $\varphi'=0$ (electron moving in medium $j= \rm o$ \emph{outside} the cylinder). We further use the identity
\begin{align}
    (k^2\eps_j + \partial_z \nabla)  \ee^{\ii q z} \xi_m(Q_j R) \ee^{\ii m \varphi} = Q_j k_j \Eb^\xi_{j,q mp}
    \nonumber
\end{align}
with $\xi_m=H_m^{(1)}$ and $j=\ii$ for $b<a$; and $\xi_m=J_m$ and $j=\rm o$ for $b>a$. This allows us to express the external field as
\begin{align}
    \Eb^\ext(\rb,\omega) = \frac{ e \pi \Qo}{ \sqrt{\epsi} c } \sum_m J_m(\Qo b ) \Eb^H_{{\rm i},q mp} (\rb)
    \nonumber
\end{align}
for the inner electron trajectory, and 
\begin{align}
    \Eb^\ext(\rb,\omega) = \frac{ e \pi \Qi}{\sqrt{\epso}c } \sum_m H^{(1)}_m(\Qi b ) \Eb^J_{{\rm o},q mp} (\rb)
    \nonumber
\end{align}
for the external trajectory. We now obtain the induced field by multiplying each of the cylindrical waves in the external field by reflection coefficients for a source placed either outside (Appendix~\ref{Sec:scattering_outside}) or inside (Appendix~\ref{Sec:scattering_inside}) the cylinder for external and inner trajectories, respectively. Inserting the result into Eq.~\eqref{Eq:EELS_dt}, the EEL probability is given by
\begin{align}
    \Gamma_\EELS(\omega) = \frac{e}{\pi \hbar \omega} \int_{-\infty}^\infty dz\, \real{\ee^{-\ii\omega z / v } E_z(b,0,z,\omega)},
    \nonumber
\end{align}
which leads to
\begin{align}
    \Gamma_\EELS(\omega) =& \frac{e^2 L}{ \hbar v^2\gamma_{\rm i}^2 \epsi} \nonumber \\ 
    \times&\sum_m 
    I_m^2\left(\frac{\omega b}{v\gamma_{\rm i}} \right)  (-1)^{m-1} \real{r_{m,pp}^\inn } \nonumber
\end{align}
for an inner trajectory ($b<a$). Likewise, we obtain
\begin{align}
    \Gamma_\EELS(\omega) =& \frac{4e^2 L}{\pi^2 \hbar v^2\gamma_{\rm o}^2 {\epso}} \nonumber \\
    \times&\sum_m  K_m^2\left(\frac{\omega b}{v\gamma_{\rm o}} \right)  (-1)^m \real{t_{m,pp}^\out} \label{eq:EELS_parallel}
\end{align}
for an outer trajectory ($b>a$). Here, we define the Lorentz factor $\gamma_j=1/\sqrt{1-(v/c)^2\eps_j}$ for an electron moving inside medium $j$. Also, we have used Eqs.~9.6.3 and 9.6.4 from Ref.~\citenum{AS1972} to simplify the expression for the EEL probability for real-valued arguments in the modified Bessel functions of the first and second kind, $I_m$ and $K_m$. The final expressions are given in terms of the reflection $r_{\sigma \sigma'}$ and transmission $t_{\sigma \sigma'}$ coefficients derived in Appendices~\ref{Sec:scattering_outside} and \ref{Sec:scattering_inside} for outer and inner trajectories, respectively.

\subsection{Long finite nanowires}

\begin{figure}
    \centering
    \includegraphics[width=0.41\textwidth]{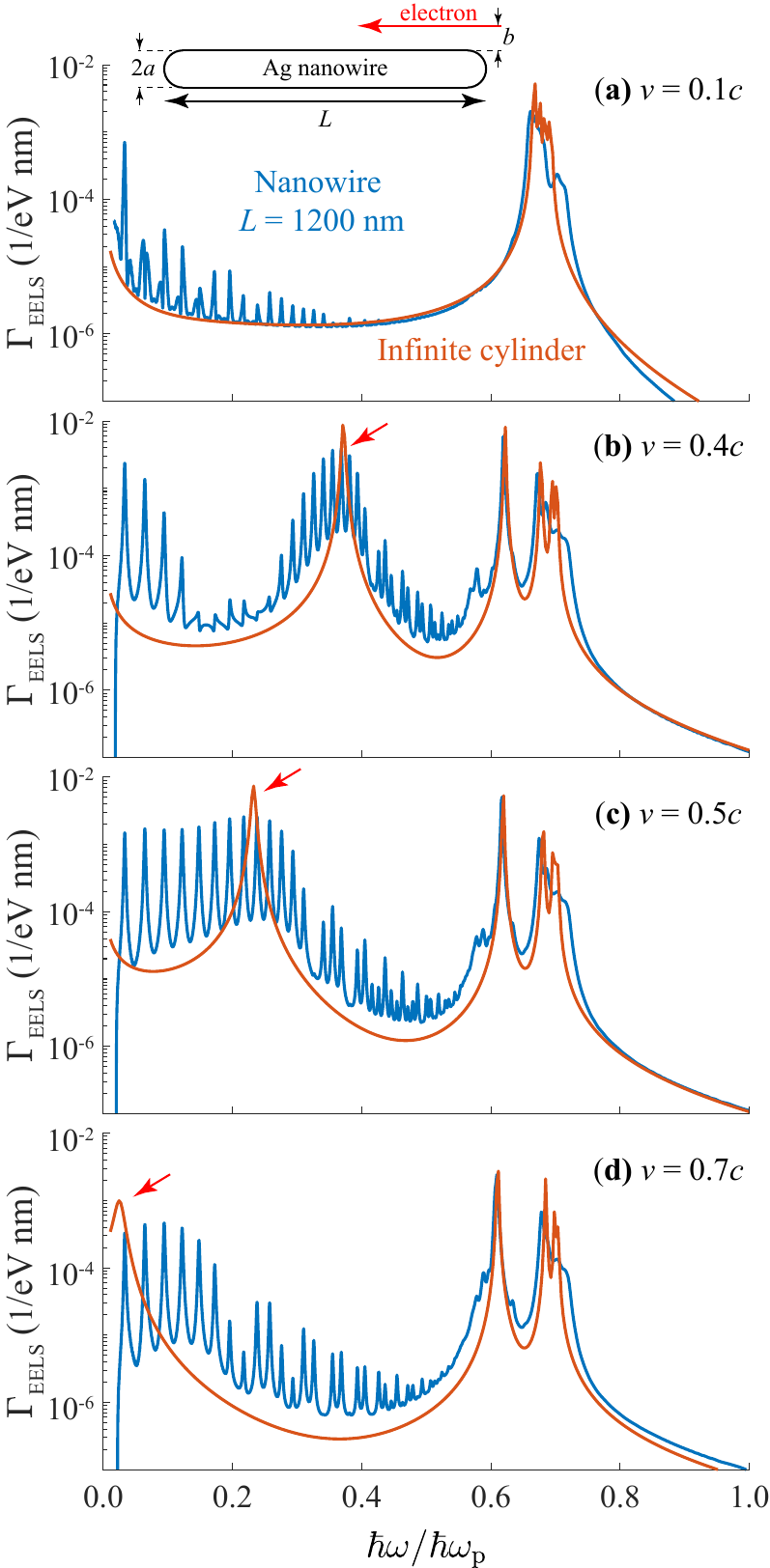}
    \caption{\textbf{EEL probability for electron trajectories parallel to a long nanowire of finite size}.
    (a-d) EEL probability for a long Drude-like silver nanowire (blue curve) of length $L=1.2$~$\um$ and radius $a=15$~nm [see inset in panel (a)] compared to the analytical solution for an infinite cylinder (red curve). We set the electron velocity to $v/c=0.1$, 0.4, 0.5, and $0.7$ (see labels). The red arrow indicates the position at which the phase velocity associated with the fundamental surface plasmon of the incident cylinder (i.e., the SPP with azimuthal number $m=0$) coincides with the electron velocity (same as in Fig.~\ref{fig:EELS_parallel}).
    }
\label{fig:fig_wire}
\end{figure}

We find it interesting to compare the results obtained for infinitely extended systems with those for long, but finite metallic nanowires, which are affected by surface plasmon reflection at the ends. Long plasmonic nanowires can be fabricated with high quality and control over geometry \cite{HGD12,paper258}. The simulation of such finite systems requires the use of numerical tools to obtain the electromagnetic fields produced by the interaction with the electron. We employ a DGTD method and apply it to a long silver nanowire \cite{SZR24} with a length $L=1.2$~$\um$ and radius $a=15$~nm. The dispersion relation in this wire is close to that of an infinite cylinder with the same radius (see Appendix~\ref{Sec:DR_infinite_cylinder}). We describe the metal through the Drude model with bulk plasma energy $\hbar\wp=9.17$~eV and damping $\hbar\gamma=21$~meV. The ends of the wire are terminated with hemispherical caps (see inset in Fig.~\ref{fig:fig_wire}a) to mimic experimental images of these types of structures \cite{paper258}.

\section{Results and discussion}

\subsection{EELS for infinitely long cylinders with perpendicular electron trajectories}

In Fig.~\ref{fig:EELS_perp}, we explore the EEL probability for perpendicular electron trajectories in cylindrical metallic structures. This is a natural configuration when studying long nanowires deposited on a TEM membrane and excited by an e-beam incident along the substrate normal. In panels Fig.~\ref{fig:EELS_perp}a and Fig.~\ref{fig:EELS_perp}b, we use a Drude-like permittivity $\epsm = \eps_0-\wp^2/\omega(\omega+\ii\gamma)$ with $\eps_0=1$, $\hbar\wp=9.17$~eV, and $\hbar\gamma=21$~meV (describing silver, although by setting $\eps_0=1$, we ignore interband transitions), and plot the momentum-resolved EEL probability obtained from Eq.~\eqref{eq:EELS_perp_q} for wire radii $a=15$~nm and $a=50$~nm, respectively, impact parameter $b-a=5$~nm, and velocity $v\approx0.55c$ ($100$~keV). Surface plasmon bands are well-resolved below the light line (white), where azimuthal modes with different $\ee^{\ii m\varphi}$ dependencies are observed. The dependence on the azimuthal number $m$ is more clearly visible for smaller radii because their corresponding surface plasmon branches are further separated in energy. We also show a frequency cut along fixed longitudinal wave vector $q_z=0.025$~nm$^{-1}$ (see insets), where we compare our analytical result with numerical solutions obtained from BEM (blue and red curves, respectively), showing an excellent mutual agreement.

We also present calculations for silver, gold, and copper wires obtained by using measured optical data \cite{JC1972} for the dielectric functions of these metals. In panels Fig.~\ref{fig:EELS_perp}c-e, we plot the momentum-resolved EEL probability for a thin cylinder of radius $a=15$~nm and impact parameter $b-a=5$~nm excited by electrons of velocity $v\approx0.55c$, exhibiting a clear surface-plasmon polariton (SPP) band. Among those noble metals, silver has the lowest intrinsic damping, thus exhibiting sharper features. Keeping constant the impact parameter $b-a=5$~nm, we study the variation with radius in the EEL probability (Fig.~\ref{fig:EELS_perp}f-h). In all three metals, there is a well-defined feature at the classical surface plasmon energy ($\hbar \omega_{\rm sp} \approx 3.7$, 2.5, and 2.3~eV for silver, gold, and copper, respectively) that becomes more pronounced when the radius is increased (i.e., approaching the limit of a planar metal surface).

Cylindrical dielectric waveguides are investigated in Fig.~\ref{fig:fig_dielectric} and shown to display features associated with waveguided modes (see illustration in Fig.~\ref{fig:fig_dielectric}a). Unlike SPPs in the metal nanowires, these modes appear both below and above the light line $\omega=q_zc$ (Fig.~\ref{fig:fig_dielectric}b). Modes within the light cone are broader because they can leak into the surrounding vacuum. Sharp trapped modes are sustained in between the light lines in vacuum and in the material ($\omega = q_z c /\sqrt{\epsi}$). This is illustrated in Fig.~\ref{fig:fig_dielectric}b for a free-standing cylinder of permittivity $\epsi=8+10^{-3}\ii$, where a small imaginary part is introduced to help visualize the modes. The influence of the permittivity is analyzed in Fig.~\ref{fig:fig_dielectric}c for $\epsi$ in the $2+10^{-3}\ii$ to $12+10^{-3}\ii$ range (i.e., from silica to silicon at optical frequencies): the stronger spatial compression of the modes as $\epsi$ is increased permeates the $q_z$-integrated spectra through the emergence of new features in the frequency region under investigation. In addition, we plot in Fig.~\ref{fig:fig_dielectric}d the ratio between the contribution to the EEL probability arising from guided modes ($\Gamma^{\rm guided}_\EELS(\omega)$, obtained by integrating outside the light cone for $|q_z|>\omega/c$) and the total EEL probability $\Gamma_\EELS(\omega)$ (integral over all $q_z$'s), illustrating the spectral range for which guided modes are the dominant electron energy-loss mechanism. For wavelengths $\lambda$ below the fundamental guided mode (i.e., $ (a/\lambda) \sqrt{\epsi-\epso} < (\alpha_0/2\pi) = 0.3827$, where $\alpha_0$ is the first zero of the $J_0$ Bessel function), guided modes amount to $>50$~\% of the total loss probability. 

\subsection{EELS for infinitely long cylinders and holes with parallel electron trajectories}

In the parallel configuration, we distinguish two possible situations where the electron moves parallel to the main axis of a cylinder or a hole. In the former, the cylinder is made of metal and the electron moves in vacuum, while in the latter, the electron moves inside a lossless material surrounded by metal (see Fig.~\ref{fig:fig1}). In Fig.~\ref{fig:EELS_parallel}a,b, we plot the EEL probability for the cylinder and hole geometries, respectively. In both cases, we take the electron to move 5~nm away from the interface (see insets) with a velocity $v\approx0.55c$ ($100$~keV). The analytical result (blue curves) compares well with full numerical simulations using BEM (red curves). Next, we use the analytical theory to apply it to compute the EEL probability for structures made of silver (Fig.~\ref{fig:EELS_parallel}c,d) and gold (panels Fig.~\ref{fig:EELS_parallel}e,f) with the metal permittivity taken from measured optical data \cite{JC1972}. We show results for different radii (see labels). For the metallic cylinder, there is a peak that shifts toward lower energies for smaller radii, corresponding to the frequency at which the phase velocity of the fundamental SPP band matches the electron velocity (i.e., at the crossing of the fundamental SPP dispersion curve and the electron line in red, see panels Fig.~\ref{fig:EELS_parallel}c,g, for a 20~nm-silver cylinder). In Fig.~\ref{fig:EELS_parallel}g, we show the transmission coefficient $\real{t_{0,pp}}$, which permits visualizing the dispersion relation of the fundamental p-polarized mode contributing to the EEL probability [see Eq.~\eqref{eq:EELS_parallel}]. Note that such a feature does not manifest in the hole configuration, for which just a high-energy feature emerges near the surface plasmon energy $\hbar\omega_{\rm sp}$. This behavior is consistent with the plasmon sum rule connecting complementary geometries (the cylinder and the hole) in the electrostatic limit (small radius and/or large $q_z$) \cite{AER96}: $\omega_{\rm cyl}^2+\omega_{\rm hole}^2=\omega_{\rm sp}^2$, where $\omega_{\rm cyl}$ and $\omega_{\rm hole}$ are the fundamental SPP frequencies of the cylinder and the hole, which are independent of $q_z$. As the radius increases, retardation effects kick in, and the sum rule is no longer satisfied.


\subsection{EELS for long finite nanowires and parallel trajectories}

To investigate finite-size effects, we adopt the DGTD method \cite{BKN11}, adapted to compute the EEL probability \cite{MNH11,SZR24} and used here to simulate a long Drude-like silver nanowire of length $L=1.2$~$\um$ and radius $a=15$~nm (see Fig.~\ref{fig:fig_wire}). For the infinitely long wire (see above), the EEL probability showed a feature at a frequency for which the fundamental SPP phase velocity coincides with the electron velocity (see red arrows in Fig.~\ref{fig:EELS_parallel}). The spectral position of such a peak is modified alongside the dispersion relation when the material, the radius, or the electron velocity are changed. For the finite cylinder, we consider in Figs.~\ref{fig:fig_wire}a-d an electron moving parallel to the cylinder, 5~nm away from the surface (blue curve), and having a velocity $v/c=0.1,\, 0.4,\, 0.5,\,$ and $0.7$. We superimpose the probability obtained for the infinite cylinder (red curves) given by Eq.~\eqref{eq:EELS_parallel}. The spectral position of the phase-matching condition is well-captured in the finite system (see below). However, as a manifestation of finite-size effects, we observe multiple peaks reaching up to the surface plasmon frequency $\hbar\wsp$ and corresponding to standing-wave resonances in the finite wire. Given the small radius of the wire, only resonances with an azimuthal number $m=0$ are observed (i.e., symmetric modes with respect to wire rotations around its axis). The mode frequencies $\omega_l$ are intuitively described by the Fabry-Pérot condition $k(\omega_l)\approx l\pi/L$ for integer values of $l$, where $q_z=k(\omega)$ is the dispersion relation of the $m=0$ fundamental mode in the infinite wire [Eq.~\eqref{eq:DR_infinite_cylinder}]. This condition is derived by setting the round-trip phase of the SPPs (after bouncing at the two ends of the wire) to a multiple of $2\pi$ (neglecting reflection phases). Obviously, the amplitude of different peaks is not constant, as those closer to the phase-matching condition (red arrows) are more prominently excited. In other words, as the EEL probability associated with a given mode is proportional to the spatial Fourier transform of the mode field for a spatial frequency $\omega/v$ \cite{paper433}, we argue that, for the modes of the long wire under consideration (which contain components with a spatial dependence $\ee^{\pm\ii k(\omega_l)z}$ along $z$), the Fourier transform (and, therefore, also the EEL probability) reaches maximum values at mode frequencies $\omega_l$ for which the phase velocity $\omega_l/k(\omega_l)$ is close to $v$.


\section{Conclusion}

In summary, we have developed a fully analytical theory to calculate the EEL probability in cylindrical geometries (individual waveguides and holes) for both parallel and perpendicular electron trajectories including retardation effects. The results are compared to numerical simulations using BEM to check their validity as well as DGTD to evaluate finite-size effects for cylinders of finite length. We consider metallic and dielectric systems, showing effects associated with the excitation of plasmonic resonances and guided modes, respectively. In the parallel configuration, excitation of the SPPs modes requires phase matching between them and the electron. For perpendicular trajectories, a wide range of longitudinal wave vectors is obtained. An extension to cathodoluminescence should be straightforward using our analytical methods. The presented compact analytical expressions are of interest to investigate cylindrical systems through EELS and can help to understand their optical characteristics.

\section*{Acknowledgments}

This work has been supported in part by the German Research Foundation (DFG) in the framework of the Collaborative Research Center 1375 “Nonlinear Optics down to Atomic Scales (NOA)” (Project No. 398816777), the European Research Council (Advanced Grant No. 101141220-QUEFES), the Spanish MICINN (PID2020-112625 GB-I00 and Severo Ochoa CEX2019-000910-S), the Catalan CERCA Program, and Fundaci\'os Cellex and Mir-Puig.

\appendix


\renewcommand{\theequation}{A\arabic{equation}}
\renewcommand{\thesection}{A}
\section{Plasmon dispersion relation of an infinite cylinder} \label{Sec:DR_infinite_cylinder}


The analytical dispersion relation of surface plasmons for an infinitely long cylinder is given by $q_z=k(\omega)$, where the frequency-dependent parallel wave vector $k(\omega)$ satisfies the equation \cite{AE1974}
\begin{align} \label{eq:DR_infinite_cylinder}
  &\nu ^2 \nu'^{2}  (\nu' \epsi \alpha_m - \nu \epso \beta_m)(\nu' \alpha_m - \nu \beta_m) \nonumber \\
  &= \frac{m^2}{a^2} (\epso - \epsi)^2 k^2 \frac{\omega ^2}{c^2}
\end{align}
with
\begin{align}
    \nu &= (k^2 - \epsi \omega^2/c^2)^{\frac{1}{2}},  & \alpha_m &= I_m (\nu a)/I'_m (\nu a),  \nonumber\\
    \nu' &= (k^2 - \epso \omega^{2}/c^2)^{\frac{1}{2}}, & \beta_m &= K_m(\nu' a) / K'_m(\nu' a). \nonumber
\end{align}
Here, $I_m$ and $K_m$ are the order-$m$ modified Bessel functions of the first and second kind, respectively, $\omega$ is the angular frequency, $k(\omega)$ is the parallel wave vector of the mode, and the primes indicates the derivative with respect to the argument.

\renewcommand{\theequation}{B\arabic{equation}}
\renewcommand{\thesection}{B}
\section{Boundary conditions for an external electron trajectory} \label{Sec:scattering_outside}


We follow the formalism introduced in Ref.~\cite{paper047} to describe the scattered fields for a source placed outside a cylinder. Then, the external field consists of a combination of $\Eb^J_{j,q_z m \sigma}(\rb,\omega)$ cylindrical waves that are scattered at the surface of the cylinder. The total electric field $\Eb(\rb,\omega)$ can be expressed as \cite{paper047}
\begin{align}
    \Eb =&\sum_{\sigma'=s,p}r_{m,\sigma'\sigma}^\out \Eb^J_{{\rm i},q_z m\sigma'}                  & & \text{(inside)}, \nonumber\\
    \Eb =&  \Eb^J_{{\rm o},q_z m \sigma} + \sum_{\sigma'=s,p}t_{m,\sigma'\sigma}^\out \Eb^H_{{\rm o},q_z m\sigma'} & & \text{(outside)}, \nonumber
\end{align}
in terms of transmission and reflection coefficients $t_{m,\sigma'\sigma}^\out$ and $r_{m,\sigma'\sigma}^\out$ for incident and scattered polarizations $\sigma$ and $\sigma'$ ($=p$ and $s$), respectively. The magnetic field is expressed as $\Hb = (1/\ii k)\nabla\times\Eb$ in terms of the electric field (Faraday's law). To determine the transmission and reflection coefficients, one needs to impose the continuity of the $\hat{\varphi}$ and $\zz$ components of the electric and magnetic fields at the cylinder surface $R=a$, leading to the matrix equations
\begin{widetext}
\begin{align}
    \left[ 
        \begin{matrix}
            r_{m,ss}^\out \\ \\ t_{m,ss}^\out \\ \\ r_{m,ps}^\out \\ \\ t_{m,ps}^\out
        \end{matrix}
    \right]
    = M^{-1}
        \left[ 
        \begin{matrix}
           \dfrac{ \Qo}{\ko}J_{m}(\Qo a) \\ \\ 
           J'_m(\Qo a)  \\ \\ 
           0 \\ \\ 
           - \dfrac{q m}{\ko \Qo a} J_{m} (\Qo a)
        \end{matrix}
    \right],
\quad \quad 
\quad \quad 
    \left[ 
        \begin{matrix}
            r_{m,sp}^\out \\ \\ t_{m,sp}^\out \\ \\ r_{m,pp}^\out \\ \\ t_{m,pp}^\out
        \end{matrix}
    \right]
    = M^{-1}
        \left[ 
        \begin{matrix}
          0  \\ \\ 
          \dfrac{q m}{\ko \Qo a} J_{m} (\Qo a)\\ \\ 
          \dfrac{ \Qo}{\ko} J_{m}(\Qo a) \\ \\ 
          J'_{m}(\Qo a) 
        \end{matrix}
    \right], \nonumber
\end{align}
with the matrix $M$ defined as
\begin{align} \label{Eq:M_matrix}
M = 
    \left[
        \begin{matrix}
            \dfrac{\xi \Qi}{\ki} J_m(\Qi a)    & \dfrac{-\Qo}{\ko} H_m^{(1)} (\Qo a)     & 0                             & 0                                      \\[8pt]
            J_m'(\Qi a)                        & -H_m^{(1) '}(\Qi a)                     & \dfrac{m q}{\ki \Qi a} J_m(\Qi a) & \dfrac{-m q}{\ko \Qo a} H_m^{(1)}  (\Qo a) \\[8pt]
            0                                & 0                                     & \dfrac{\Qi}{\ki} J_m(\Qi a)     & \dfrac{-\Qo}{\ko} H_m^{(1)} (\Qo a)      \\[8pt]
            \dfrac{\xi m q}{\ki \Qi a}J_m(\Qi a) & \dfrac{-m q}{\ko \Qo a} H_m^{(1)} (\Qo a) & \xi J_m'(\Qi a)                 & -H_m^{(1) '} (\Qo a) 
         \end{matrix}
    \right].
\end{align}

\renewcommand{\theequation}{C\arabic{equation}}
\renewcommand{\thesection}{C}
\section{Boundary conditions for inner electron trajectories} \label{Sec:scattering_inside}

For a source placed inside the cylinder, the external field (for $R<a$) can be expanded in terms of outgoing cylindrical waves $\Eb^H_{{\rm i},q_z m \sigma}(\rb,\omega)$ near the surface. The total field $\Eb(\rb,\omega)$, including surface scattering, then becomes \cite{paper385}
\begin{align}
    \Eb =&  \Eb^H_{{\rm o},q_z m \sigma} + \sum_{\sigma'=s,p}r_{m,\sigma'\sigma}^\inn \Eb^J_{{\rm i},q_z m\sigma'}                  & & \text{(inside)}, \nonumber\\
    \Eb =&  \sum_{\sigma'=s,p}t_{m,\sigma'\sigma}^\inn\Eb^H_{{\rm o},q_z m\sigma'}   & & \text{(outside)} \nonumber
\end{align}
in terms of coefficients $t_{m,\sigma'\sigma}^\inn$ and $r_{m,\sigma'\sigma}^\inn$. The latter are found by imposing the boundary conditions, which lead to the matrix equations
\begin{align}
    \left[ 
        \begin{matrix}
            r_{m,ss}^\inn \\ \\ t_{m,ss}^\inn \\ \\ r_{m,ps}^\inn \\ \\ t_{m,ps}^\inn
        \end{matrix}
    \right]
    = M^{-1}
        \left[ 
        \begin{matrix}
           -\xi \dfrac{ \Qi}{\ki}H^{(1)}_{m}(\Qi a) \\ \\ 
            -H^{(1)'}_{m}(\Qi a) \\ \\ 
           0 \\ \\ 
           -\xi \dfrac{q m}{\ki \Qi a} H^{(1)}_{m} (\Qi a)
        \end{matrix}
    \right],
\quad \quad
\quad \quad
    \left[ 
        \begin{matrix}
            r_{m,sp}^\inn \\ \\ t_{m,sp}^\inn \\ \\ r_{m,pp}^\inn \\ \\ t_{m,pp}^\inn
        \end{matrix}
    \right]
    = M^{-1}
        \left[ 
        \begin{matrix}
          0  \\ \\ 
          -\dfrac{q m}{\ki \Qi a} H^{(1)}_{m} (\Qi a)\\ \\ 
          -\dfrac{ \Qi}{\ki}H^{(1)}_{m}(\Qi a) \\ \\ 
          -\xi H^{(1)'}_{m}(\Qi a) 
        \end{matrix}
    \right], \nonumber
\end{align}
where $M$ is the matrix defined in Eq.~(\ref{Eq:M_matrix}).
\end{widetext}


%

\end{document}

%% file: defs.tex
\usepackage{commath} 

\def\ii{{\rm i}}  \def\ee{{\rm e}}

\newcommand{\real}[1] {\mathopen{}{\rm Re}\left\{#1\right\}\mathclose{}}

\def\Eb{{\bf E}}

\def\Hb{{\bf H}}  
  
\def\Jb{{\bf J}}  
  \def\kb{{\bf k}}

\def\Qb{{\bf Q}}  \def\qb{{\bf q}}
\def\Rb{{\bf R}}  \def\rb{{\bf r}}

  \def\vb{{\bf v}}

\def\xx{\hat{\bf x}}   \def\vv{\hat{\bf v}} 
\def\zz{\hat{\bf z}}  
\def\0b{{\bf 0}}

\def\wp{\omega_{\rm p}}      \def\wsp{{\omega_{\rm sp}}}

\def\eps{\epsilon}    
\def\epsm{\epsilon_{\rm m}}

\def\epsi{\epsilon_{\rm i}}
\def\epso{\epsilon_{\rm o}}

\def\eps1{\epsilon_{\rm 1}}
\def\Qi{Q_{\rm i}} \def\ki{k_{\rm i}}
\def\Qo{Q_{\rm o}} \def\ko{k_{\rm o}}

\def\um{{\mathrm{\text\textmu m}}}

\def\inn{{\rm in}} \def\out{{\rm out}}

\def\EELS{{\rm EELS}}

\def\ko{{k_{\rm o}}}      \def\kko{{k^2_{\rm o}}}
 \def\QQo{{Q^2_{\rm o}}}
       
\def\eps{\epsilon}        \def\inn{{\rm in}} \def\out{{\rm out}}

\def\ext{{\rm ext}}        \def\ind{{\rm ind}}

